# Development of a Digital Front-End for Electrooculography Circuits to Facilitate Digital Communication in Individuals with Communicative and Motor Disabilities

Andre Heid Rocha da Costa, Keiran Robert O'Keeffe

October 1, 2024

**Abstract:**

This project developed a cost-effective, digital-viable front-end for electrooculography (EOG) circuits aimed at enabling communication for individuals with Locked-in Syndrome (LIS) and Amyotrophic Lateral Sclerosis (ALS). Using the TL072 operational amplifier, the system amplifies weak EOG signals and processes them through an Arduino Uno for real-time monitoring. The circuit includes preamplification, filtering between 0.1 Hz and 30 Hz, and final amplification stages, achieving accurate eye movement tracking with a 256 Hz sampling rate. The approach to this was described in detail, with a comparison drawn between the theoretical expectations of our circuit design and its viability in contrast to the actual values measured. Our readings aimed to create an interface that optimized max-gaze angle readings by outputting a maximum reading at values above the baseline theory of our amplification circuit. From this, we measured the latency between the serial output and action, analyzing video recordings of such readings. The Latency value read reached around $4 \cdot 10^{-4}$ seconds, which is within the tolerance for proper communication and did not seriously affect the readings. Beyond this, high-resolution signal fidelity was achieved (with an SNR of 38.65 dB). However, some interference and noise-originated limitations remain despite achieving reliable signal amplification. Its limitations mean that future improvements will focus on reducing environmental interference, optimizing electrode placement, applying a novel detection algorithm to optimize communication applications, and enhancing signal clarity to make the system more effective for real-world applications. Even given that in its current iteration, it still encapsulates a viable design.

**Table of Contents:**

## Introduction

Tetraplegia, particularly in individuals with conditions like Locked-in Syndrome (LIS) and Amyotrophic Lateral Sclerosis (ALS), severely impairs the ability to communicate due to the near-total paralysis of voluntary muscles.[1] These conditions create significant barriers to interaction with traditional communication devices,[2] leaving eye-tracking technology as one of the few viable methods for facilitating communication. Eye-tracking, which leverages individuals' intact brain function and residual eye movement, offers a critical solution to bridging communication gaps for those with limited motor functions. Due to the nature of conditions like LIS, which results from damage to the pons region of the brain and often stems from traumatic injuries, strokes, or neurodegenerative conditions, this method is one of the few viable communication mediums to facilitate and increase the quality of life of individuals cost-effectively afflicted by this condition.[3] The prevalence of these conditions is likely underreported due to the absence of consistent tracking, especially in developing countries. The central challenge faced by those with LIS and ALS is not only the physical impairment but also the inability to communicate effectively, further isolating them from society and complicating medical care.

Current solutions to this communication barrier often need to be revised regarding high costs, complexity, and limited usability. These factors make such systems inaccessible to many, especially in resource-limited environments, communities, and countries, raising concerns about the democratization of information and the engagement of these within civil society. Given this difficulty in accessibility and the lack of thorough research regarding these issues, this paper aims to address these barriers of entry and study the feasibility of a novel, electrode-mediated front-end utilizing EOG technology to create an interface that mediates communication between a computer interface and the outside world. This study evaluates the specifications of implementing a system to facilitate communication for individuals affected by conditions such as LIS and ALS. The objective is to develop a prototype with efficient signal readability and minimal latency, utilizing cost-effective communication technologies. The project will encompass a literature review of viable circuit designs and alternatives, theoretical analysis, circuit design, and practical testing to substantiate the potential of these communication solutions for the target user group.

---

[1] Cleveland Clinic. (2022). *Locked-in Syndrome (LiS): What It Is, Causes & Symptoms*. Cleveland Clinic. https://my.clevelandclinic.org/health/diseases/22462-locked-in-syndrome-lis

[2] Voity, K., Lopez, T., Chan, J. P., & Greenwald, B. D. (2024). Update on How to Approach a Patient with Locked-In Syndrome and Their Communication Ability. *Brain Sciences*, *14*(1), 92–92. https://doi.org/10.3390/brainsci14010092

[3] Divya Swami Nathan, Vinod, A. P., & Thomas, K. P. (2017). *An electrooculogram based assistive communication system with improved speed and accuracy using multi-directional eye movements*. International Conference on Telecommunications and Signal Processing. https://www.semanticscholar.org/paper/An-electrooculogram-based-assistive-communication-Nathan-Vinod/41b8ad825117d9424822e76a805393a699d8ca12

## Literature review

### Overview of Eye-Tracking Technology

Some eye-tracking systems rely on infrared cameras or video-based tracking, which, while effective, are costly and require specialized hardware.[4] These systems often lack portability and accessibility, limiting their widespread use, especially in low-resource settings. Given this, electrodes will be applied to a physical medium to facilitate the reading of signals and motion. Electrode systems based on biopotential electrodes, such as the interface planned out in this paper, act as a medium between biological tissue and electronic measuring circuits, which in this case would be an amplifier circuit. The electrodes perform a transduction process of the ion current displayed through the eye into the electronic current that may be read and transported into a circuit responsible for reading and detection.[5] The electrode placement in this configuration will optimize the readability of signals throughout the eye and its precise motion. This technology is called electrooculography, or EOG.

---

[4] Brousseau, B., Rose, J., & Eizenman, M. (2018, November). *SmartEye: An Accurate Infrared Eye Tracking System for Smartphones*. ResearchGate; unknown. https://www.researchgate.net/publication/335195036_SmartEye_An_Accurate_Infrared_Eye_Tracking_System_for_Smartphones

[5] Dincer, C., Bruch, R., Guder, F., & Rama, E. (2019). *Electrochemical signal transduction. There are four main types of...* ResearchGate. https://www.researchgate.net/figure/Electrochemical-signal-transduction-There-are-four-main-types-of-electrochemical-methods_fig3_333132057

At the heart of electrooculography (EOG) is detecting the corneal-retinal potential (CRP),[6] where the cornea holds a positive charge and the retina a negative charge. The anterior eye placement, the cornea, is slightly positively charged. In contrast, the posterior section of the eyeball is negatively charged.

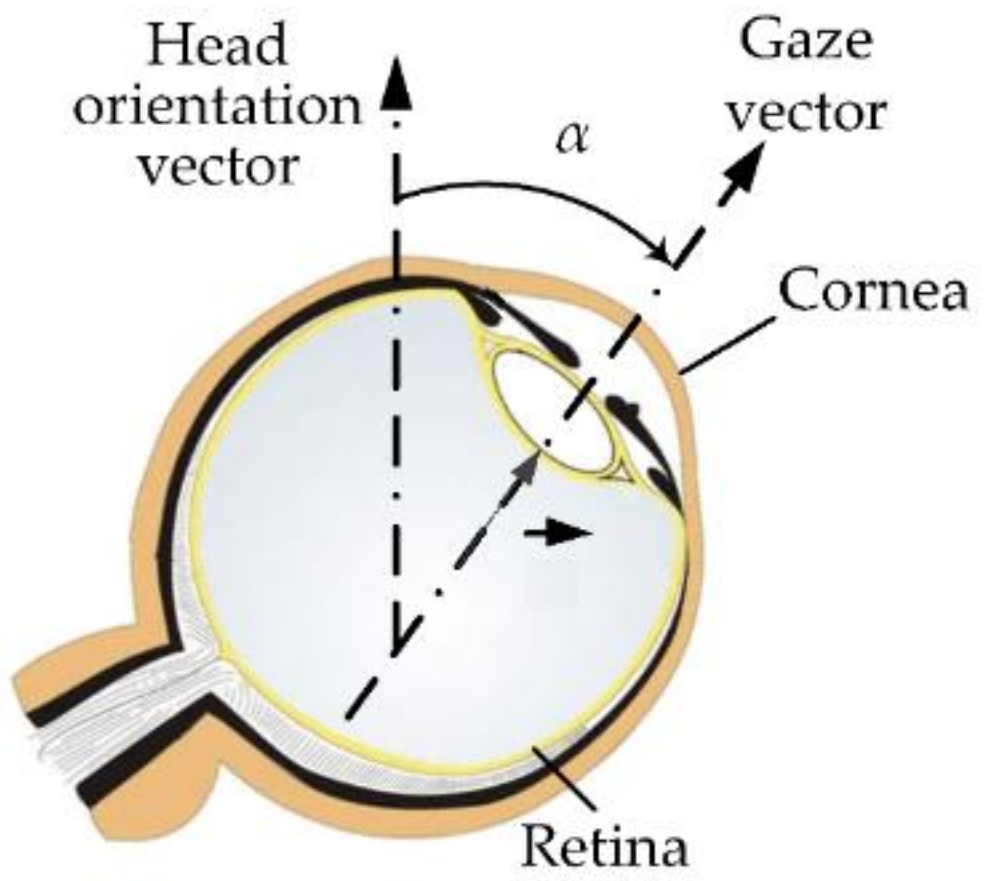

*Figure 1: Vector gaze angle orientation, Lopez et al. 2020, p.2*

Thus, eye motion can be modeled as a vector, in which the displacement of the eye vertically or horizontally changes a fixed head orientation vector, normal to the placement of gaze at an initial position, about a gaze vector, perpendicular to the position of the eye (fig. 1). Electrode placement in this configuration is responsible for measuring and detecting that change in potential as the dipole vector changes with the angular displacement of the eye. Electrode placement is at the heart of many of the issues targeted in this paper, so the specific placement details will be explored, as will the efficiency of this positioning.

---

[6] Creel, D. J. (2019). The electrooculogram. *Handbook of Clinical Neurology*, *160*, 495–499. https://doi.org/10.1016/B978-0-444-64032-1.00033-3

***Electrode Placement***

Electrode placement is essential to capture an accurate reading, and its contributions to signal acquisition are on par with some of the crucial experimental configurations to maximize readings and optimize the production of high-quality, readable signals. Electrode placement is dependent on the readings required. For vertical displacement, a positive channel electrode should be placed 1.0 cm above and below the eye, typically slightly above that benchmark due to the presence of the eyebrow, while the other is unaltered.

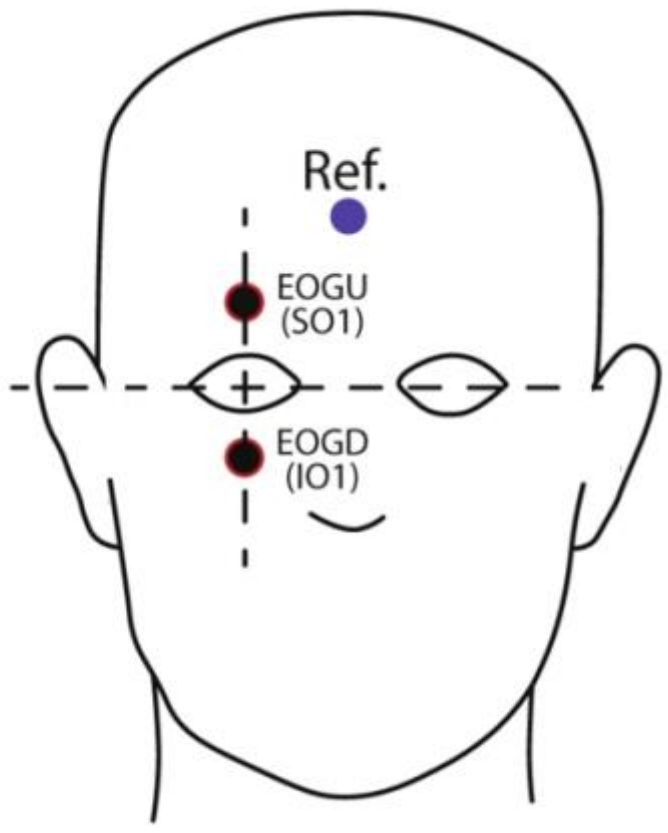

*Figure 2: Eog position for vertical displacement detection and reference point position*

However, this shouldn't modify the results of the tests done. Although not targeted in this paper, horizontal motion is a possible extension of this experiment; it is responsible for measuring horizontal displacements in potential difference and, as such, is also based on specific positions. For the horizontal electrode placement, the positive and negative channels may be placed interchangeably, and they are specifically meant to be 1.0 cm from the right outer canthus (ROC) and 1.0 cm from the left outer canthus (LOC). [7] This positioning is essential for possible extensions to this research paper. It's important to note that a reference electrode is also positioned in a neutral head area, such as the forehead or the mastoid, for calibration issues and the proper functioning of the amplifier circuit's readability. For this specific paper, the neutral used for reference is the forehead. This electrode configuration is essential for EOG technology.

[7] Belkhiria, C., & Peysakhovich, V. (2020). *FIGURE 1 | Six different electrode positions, used to measure EOG. The...* ResearchGate. https://www.researchgate.net/figure/Six-different-electrode-positions-used-to-measure-EOG-The-blue-circles-indicate-the_fig1_347848649

**Theoretical baseline as a basis for research**

During the literature analysis portion of this research, much time was spent organizing and attempting to interpret signal readings in a concise model, which could make the readings of the signal easily interpreted. This is based on research on prior EOG circuits, which have given a baseline comparison for our readings on which we can base our interpretation.

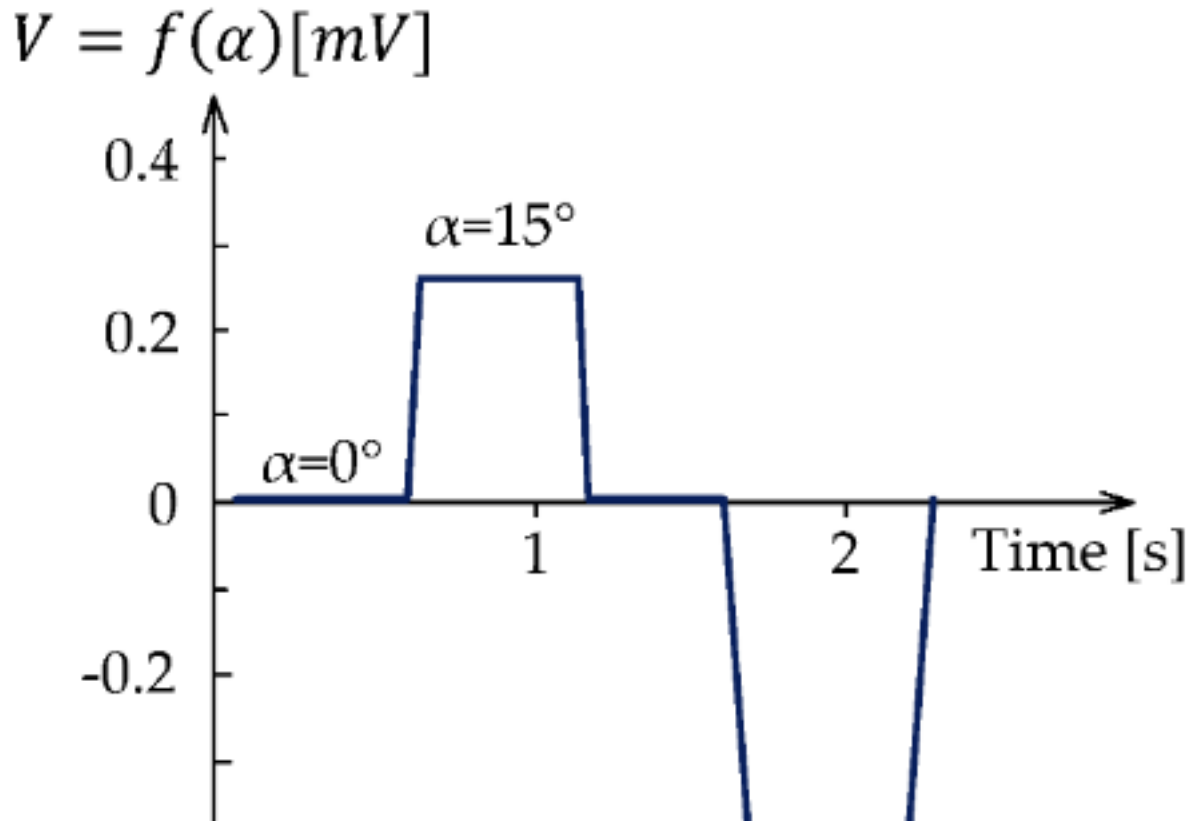

*Figure 3: : Theoretical reading values for analog detection. Digital would only measure the max 30-degree displacement, Lopez et al. 2020, p.*

Such a reading (fig. 3) can be noted as a function of gaze angle, the maximum angle displacement of your eye about the gaze vector, and the voltage output of the eye motion recording proportional to the angle being read. Mathematically, the only complexity would come from finding an angle $f(\alpha)$ that can be easily interpreted and would account for the total input and output of the circuit. Before constructing a tangible circuit, we parted from this direction and attempted to model the circuit around it. What was done, therefore, was to create an experimental setup in which the only readings being interpreted and analyzed were the maximum angular displacement measurements. So, the only readings accumulated for this round of experimental tests would be the achievable maximum voltage readings. To estimate this, it was essential to analyze and research the specific details about the TL072 and address possible sources of error and mistakes in our reading.

## Theoretical functionality and viability

Our guiding process throughout this study was to analyze the viability of an operational amplifier circuit that would interpret channel input from electrodes to produce a viable EOG output reading. For the extent of this paper, the readings will be analyzed through serial output in the form of a digital medium, in which readings above a theoretical upper baseline output a positive reading. In contrast, those below this baseline will output a 0. Conventionally, this may need to be more practical. However, given our current resources, the limited extent to which productive structure analysis may be created restricts our ability to design and interpret a wider availability of single processes at varying gaze vector positions. Regardless, a singular circuit design will be followed based on the literature principles during the literature review section. These are valuable in setting the baseline readings around which our amplification circuit will be built.

**Pre-amplifier**

Given the weak output of the electrodes and the slight potential accrued by the gaze angle function, the signal must be processed in a pre-amplifier configuration before being amplified further to be refined and processed more efficiently. The use of an instrumentation amplifier (INA) is optimal in this situation, given its high input impedance, minimizing loading on the biological source, as well as a high common-mode rejection ratio (CMRR) whose sole purpose is the elimination of a specific noise range from AC interference in the order of 60 Hz (due to geographical location, this is the AC interference dealt with in this study).[8] These are also 6 to 60 times the frequency range studied for EOG circuits, which range from 0.1 Hz to 10 Hz. By accounting for electromagnetic interference early, further processing will require less focus on noise mitigation.

The preamplifier configuration in this setup was created utilizing a 3-Op amp instrumentation amplifier. The amplifier selected throughout this work is the TL072, a JFET-input op-amp, meaning the only configuration possible in which we can minimize common mode rejection (CMR) utilizes a FET-configured 3-op-amp system instrumentation amplifier. Although this setup does not maintain the same optimization benefits of utilizing bipolar amplifiers, which can retain higher CMR and lower input offset voltage drift,[9] it is still optimal given the gain being worked with and the general intent of the circuit. The signal amplified through the two electrode channels is interpreted and sent forward towards a non-inverting amplifier setup, which, for the intents of the theoretical design, will serve as a band-pass filter for the signal outputted thus far.

---

[8] Ocanto, P. (2024). *1964 WEC Electrical Transmission And Distribution Reference Book 4th Ed.pdf*. Scribd. https://www.scribd.com/document/630826188/1964-WEC-Electrical-Transmission-And-Distribution-Reference-Book-4th-Ed-pdf

**Band-pass filter**

EOG signals have a specific frequency range that must be isolated from external noise, for example, any muscle twitching or head displacement that functions at different frequencies than those of the retinal-cornea displacement measurements taking place through the ocular electrodes. The bandpass filter, therefore, filters out those signals and isolates the targeted frequency. This configuration should be in the low and high-cut marks that are equivalent to the measurements of the EOG. For this reason, low-pass filtering, and high-pass filtering are combined into a bandpass filter, which enables the isolation of specific frequency bands as specified prior.

The purpose of the high- and low-pass filters is a widely utilized value whose diminishing characteristics for low- and high-frequency noise preserve the specific frequency range targeted as part of this circuit. To understand the values and limitations of this particular function, it is essential to break down the two ranges to measure and weigh the values and limitations that must be accounted for in constructing the circuit.

***Low-pass filtering***

It diminishes high-frequency noise and artifacts and preserves the lower-frequency elements of the signal. This functionality is intended to protect the pertinent measurements of eye movement analysis, which may also enhance the measured signal-to-noise ratio. However, the potential to introduce phase distortion results in a time delay for the filtered signal, which must be accounted for to establish a specific communication methodology for EOG functions.

[9] Jaquay, J. W. (1977). Designers Guide to: Instrumentation amplifiers. *Experimental Techniques*, *2*(2), 40–43. https://doi.org/10.1111/j.1747-1567.1973.tb01631.x

***High-pass filtering***

On the other hand, a high pass filter is responsible for eliminating low-frequency noise, mainly in the form of baseline and low-frequency drift, which originated from lower-frequency noise sources, such as drift from the electrodes. One advantage of this system is enhancing the eye-movement frequency components within the specified range of values targeted. This eliminates wander from possible experimental set-up flaws and controls for a few difficult-to-account-for variables, such as electrode drift. On the other hand, the loss of valuable info within the low-frequency component may warrant significant concern, given that information for possible analog interpretations of signals is lost in the analysis section. A potential mitigation for these would be establishing a digital potentiometer to calibrate the High-pass filter configuration individually. Combining both of these, the benefits of both the low and high-pass filters can enable proper isolation of specific frequency bands relevant to the analysis of eye movements. A better way of accounting for possible restrictions to band-pass filtering would be incorporating adaptive filtering techniques, which, if accounting for computational complexity, may continuously update the filter coefficients based on the signal input.

To theoretically verify the band-pass filter's functionality, the following frequency response graph was computed on proteus, a standard circuit design and simulation software, to make a proper signal range estimation. From this, it is plausible to state that the readings for both the high and low pass filters are optimal and that the functionality of the instrumentation amplifier in configuring this setup was viable. Such reading can be, therefore, used in a real experimental circuit design and serve as a functional alternative to the EOG readings required.

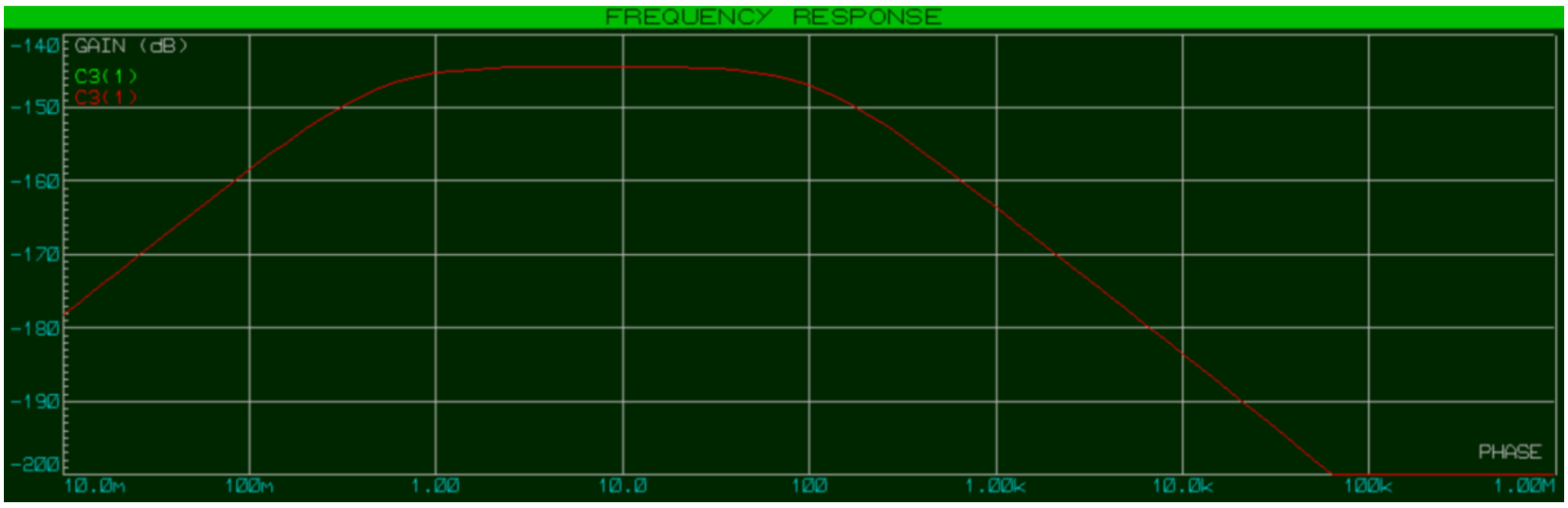

*Figure 4: Theoretical band pass filter reading in Proteus*

**Amplifier circuit**

After reducing the noise in the filtering stage, the signal might become too weak for further processing or conversion to digital format. To address this, an additional amplification stage is necessary. This stage usually involves a non-inverting amplifier with a gain that depends on the specific amplification needs. This amplifier stage ensures that the readings fall within the appropriate range for a digital-analog converter, especially when dealing with high-voltage ranges. The non-inverting operational amplifier was chosen for its ease of use with a TL072. The selected chip has high input impedance, low total harmonic distortion, wide slew rate, and low input bias current. These are crucial for achieving a high-fidelity signal, low noise floor, and reducing the signal-to-noise ratio. Most important to this application, the TL072 is one of the best viable alternatives, possesses an excellent price-to-performance ratio, is widely available, and has an inexpensive architecture.

**Driven-right leg**

Although the simple selection of a JFET operational amplifier like the TL072, common-mode interference may still be present due to the nature of the human body and its role as a sort of “antennae” in capturing and furthering possible noise in the circuit. The specific operation of this circuit is designed to actively cancel common-mode interference by injecting a compensating signal back into the body. This is crucial for highly sensitive bioelectric signals, like EEG or EOGs. By connecting a third reference electrode to a neutral reading near the target channels, we can account for significant issues arising from this circuit. Lack of isolation between the circuit and the body may warrant considerable current flowing through the third electrode. The electrode must be connected directly to this circuit to account for this. It fundamentally reduces the effective electrode resistance by several orders of magnitude while also allowing only a safe amount of current to flow through the third electrode, which is only sufficient for the signal to be reduced.

## Equipment and Methodology

The methodology employed in this study follows a comprehensive process aimed at designing, simulating, and testing an effective electrooculography (EOG) circuit, which interfaces with a microcontroller for digital signal acquisition. The core of the EOG circuit revolves around the **TL072 operational amplifier (op-amp),** selected due to its high input impedance $(\sim 10^{12}\Omega)$[10] and low noise characteristics, which are vital for amplifying weak EOG signals ranging between $50\ \mu V$ and $3500\ \mu V$.[11] The role of this is to incorporate the theoretical functionality of the circuit measured prior and check its viability for this specific application.

The first stage in the circuit involves a preamplifier, constructed using an instrumentation amplifier configuration, established with a gain of 40000 V/V to bring the signals within a detectable range. This amplification is governed by the equation Gain = $A_v = \left(1 + \frac{2 \times R_2}{R_{gain}}\right) \times \left(\frac{R_4}{R_3}\right)$, which, for the sake of the calculations applied in this circuit design, around $40000\frac{V}{V}$ , where the R values used are a function of the theoretical guidelines presented further down. After preamplification, the expected output voltage is 5 mV to 350 mV, sufficient to pass through the filtering stages for further processing.

---

[10] Texas Instruments. (2023). *TL07xx Low-Noise FET-Input Operational Amplifiers*. https://www.ti.com/lit/ds/symlink/tl072.pdf

[11] Donnelly, D., Hofflich, B., Lee, I., Lunardhi, A., & Tor, A. (n.d.). *Electrooculography (EOG) Eye Gaze Communication Device*. Retrieved September 18, 2024, from https://isn.ucsd.edu/courses/beng186b/project/2021/Donnelly_Hofflich_Lee_Lunardhi_Tor_EOG_eye_gaze_communication_device.pdf

A band-pass filter was designed using the aforementioned methodology, meant to filter out unwanted noise and ensure that only the relevant EOG signal frequencies between 0.1 Hz and 30 Hz are retained. The filter consists of a high-pass component with a cut-off frequency of 0.5 Hz to eliminate low-frequency noise, such as baseline drift, and a low-pass component with a cut-off frequency of 30 Hz to remove higher-frequency interference.

Following the filtering stage, the signal is passed through a second amplification stage to scale the voltage for microcontroller compatibility. To meet this requirement, the second amplifier stage, also based on the TL072, was configured with a gain of $10000\ \frac{V}{V}$, based on the theoretical approach of the equation $A_v = 1 + \frac{R_2}{R_1}$ . This resulted in a theoretical total system gain of 50000 V/V, with the amplified signal ranging from 0.5 V to 3.5 V, making it suitable for the Arduino Uno's ADC. The Arduino Uno, with its 10-bit ADC resolution,[12] converts the amplified signals into digital data, providing a fine granularity of approximately 5/1024 V per bit for measuring small voltage changes corresponding to eye movements. The data acquisition process is critical for interpreting the EOG signals, and the Arduino Uno was configured to sample the signals at 256 Hz, which is more than sufficient for capturing the 0.1 Hz to 30 Hz frequency range of the EOG signals. The digital output from the Arduino is monitored through the serial interface, where it is displayed as voltage vs. time to track the EOG signals in real time. This provides the basis for identifying the angular displacement of eye movements based on the voltage changes. The best possible gain value measure for experimental testing as around 50001 V/V, given the difficulty of approximating the theoretical amounts to those obtained experimentally.

---

[12] Itterheimová, P., Foret, F., & Kubáň, P. (2021). High-resolution Arduino-based data acquisition devices for microscale separation systems. *Analytica Chimica Acta*, *1153*, 338294. https://doi.org/10.1016/j.aca.2021.338294

Throughout this prototying stage, as we were incorporating the relevant literature for this specific applicability range, challenges arose regarding the actual cut off values for the specific band-pass filter readings that'd become a significant deterrant to plausible sources of external noise and interference. AS a consequence, much of our time was spent calculating a high-fidelity signal circuit design which would incorporate the complexity of possible sources of error and mitigate the overall impedance, noise and instability that arose as a result of our unoptimized laboratory configuration. Plausible approaches to this problem would focus on accounting for conditions in the laboratory in which the signal is being tested, as well as fidning achievable models that may provide a way of accounting for low input impedance readings, instability in functionality and use, as well as noise and interference, by far the single most significant deterrant to an accurate signal reading.

## Technical Functionality of EOG and Amplifier Circuits

Based on this information and the methodological baseline presented, the following circuit diagram could be devised:

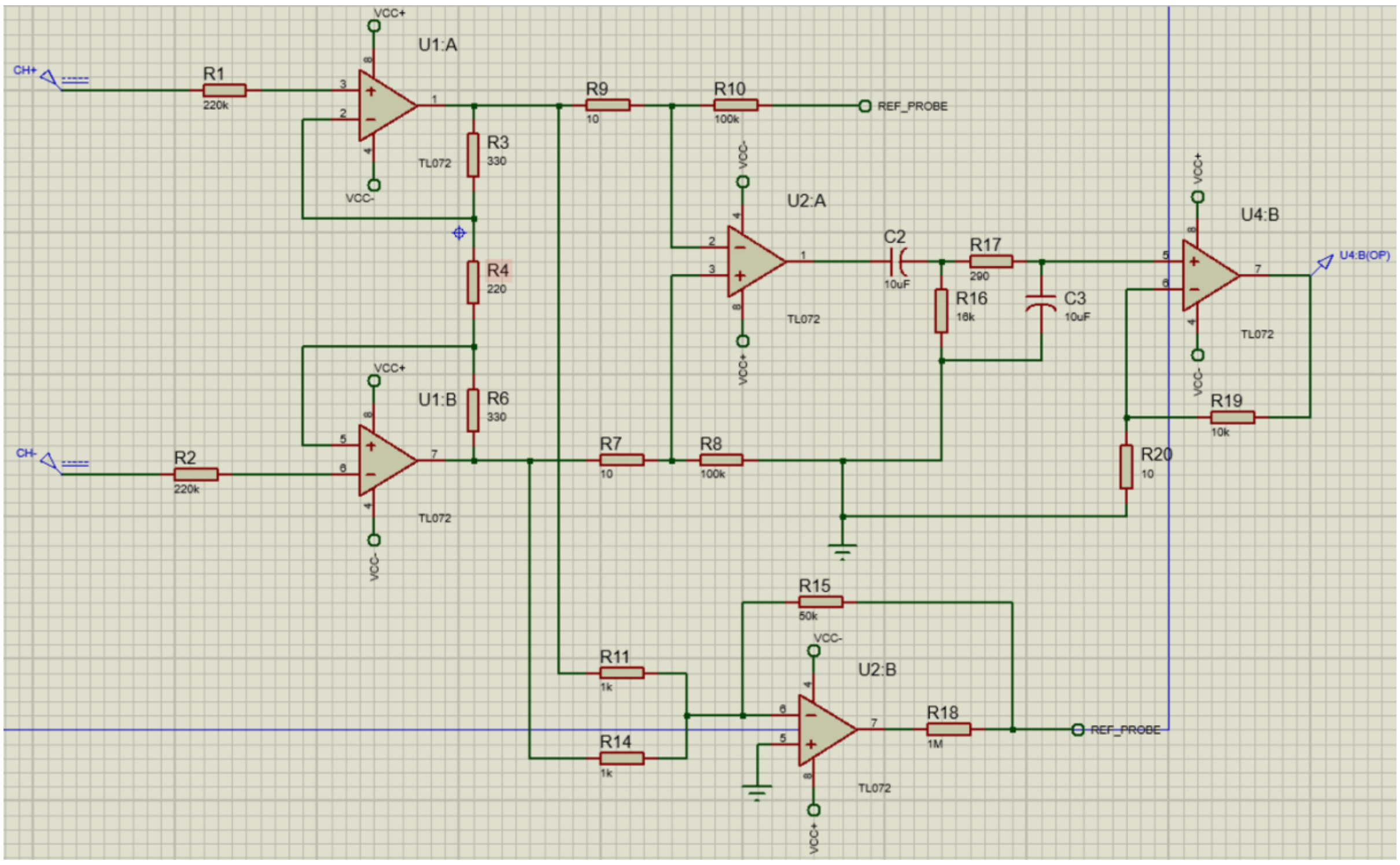

*Figure 5: Circuit diagram constructed on Proteus*

This diagram, constructed in proteus, adheres to the principles of EOG circuit design and the specific phases chosen to amplify the circuit. Due to the lack of high-impedance equipment and a way of preventing the possible intervention of external interference and noise, it was essential to adhere as closely as possible to the circuit guidelines presented earlier, with a pre-amplifier, band-pass filter and amplifier stage. From the data sheet available for this circuit reading, the total gain computed would amount to the following:

$$A_v(instrumental) = \left(1 + \frac{2\ \times R_2}{R_{gain}}\right) \times \left(\frac{R_4}{R_3}\right) = \left(1 + \frac{2 \times 330\Omega}{220\Omega}\right) \times \left(\frac{100000\Omega}{10\Omega}\right)$$

Readings would pass through the circuit, be amplified, and read through a CSV file outputted by an Arduino serializer. We noted the difficulties in creating tangible disparate readings reliant only on the outputted voltages to interpret those data points. As such, the different path was to pass all signals through an analog-to-digital converter. The signals were processed and interpreted through a Python program that measured the maximum peaks detected for the lower-range and upper-range limits. These are the upper and lower gaze angle readings at positive and negative gaze vector displacement. Once accounted for, an upper boundary would be determined to ease the possible output reading in the graph. In contrast, all other readings were reduced to a specific upper and lower range to analyze the potential interference of noise in the detection of signals.

This information convinced us that the method approach would be viable for testing. After incorporating a signal attenuation sub-circuit in the input channel to reduce the possible voltage from returning into the readings through unidirectional diodes, the circuit was soldered onto a proper protoboard and was ready for testing. Data collected from volunteers yielded mitigating factors that had to be accounted for, and a total of 5 trial runs were collected. From this, a counter study was done at each data point to interpret the tangibility of the results due to possible sources of error from human sources, movement, capacitive coupling in the readings, and other sources. As a result, only two select readings were deemed optimal, and such a graph was depicted as a voltage graph by time, where the voltage was adjusted for readability utilizing the digital-to-analog principle we set out to describe in prior stages.

The results were shown as follows:

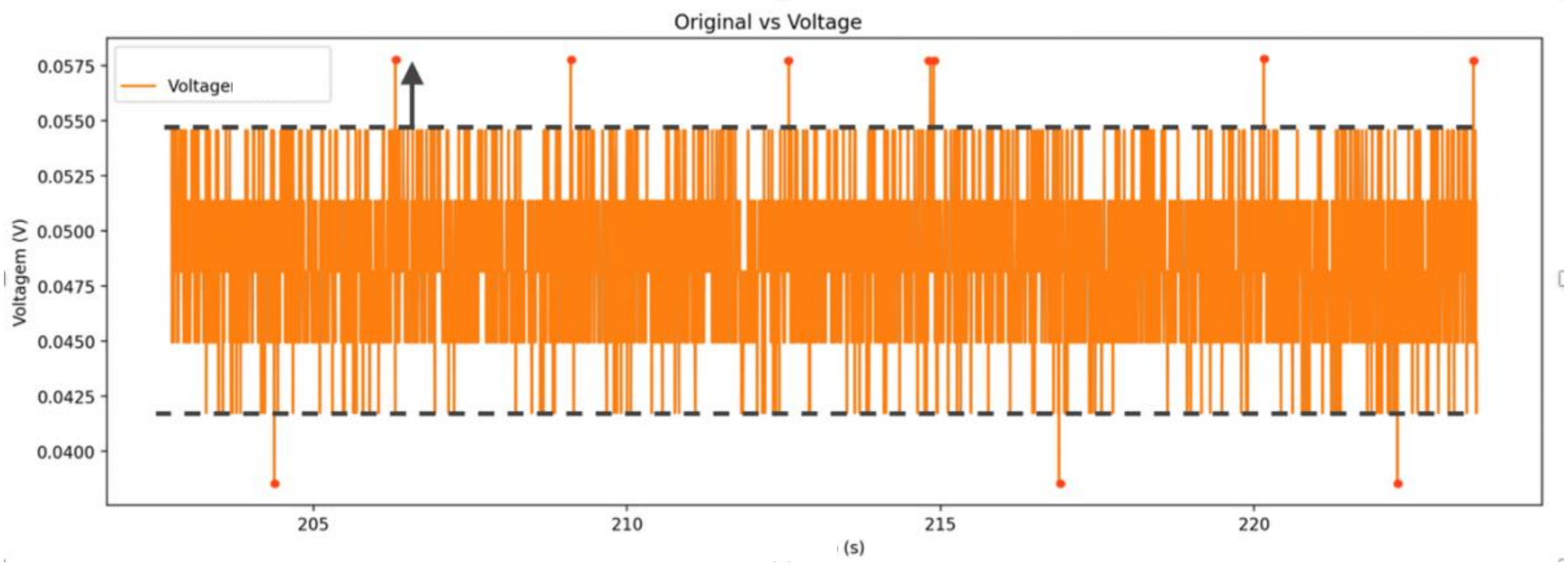

*Figure 6: Graph of resulting readings*

The interpretation of this data leads us to believe that there are still significant sources of error and readings that may yield credence to the idea that the interference generated extends far beyond the circuit design itself. The reduction of these will be paramount to addressing the significance of this work for mediating communication among individuals with LIS and ALS. However, from what was gathered, it is still evident that a significant range of readings portray achievable and well-measured readings. Every peak in red in the figure describes a maximum ocular displacement, showing that the linear relationship between gaze angle and voltage readings is being captured by the electrodes and adequately measured by the Arduino. Concurrent literature corroborates this claim, with the maximum reading of 0.0550 V being consistent with prior readings of maximum displacement readings for EOG circuits that include an amplifier setup. The opposite gaze displacement, now positioned downwards, is also captured in this graph, with readings below a 0.0400 V threshold. All are placed comfortably within the range of values represented by the grey dotted line.

***Latency & Signal to noise ratio***

In order to find a reasonable estimate of the latency of the circuit in question, the following equation was used with the mean readings computed from each trial run.

$$\text{Mean Latency} = \frac{\sum(\text{time differences})}{N} = 0.00416\,\text{s}$$

This indicates a common mean difference in the written output readings for this specific circuit design. It suggests a variation in the total readings N and the time interval measured between these. However, although the latency of the circuit is considerably low and provides an optimal interpreted value that is significant to the final goals pursued in this research article, it's important to note the experimental downsides to the prior methods used. As part of the original measurement of signal accuracy in the graph depicted earlier, human-dependent variables were used, such as comparison to recordings and dependence on visual observation. However, this was seen as inaccurate and prone primarily to human error. The objective computation of latency, a fundamental feature of this circuit's viability analysis, must be done using raw output data from the serializer. This already accounts for the millisecond delay from the serializer reading as packet loss becomes a concern in overall reading tenacity and, as such, provides a reading accurate enough to be considered significant. In addition, we can also ensure the results by stating the qualitative deduction of accuracy from prior graph interpretation and corroborating the significance of their readings through these more qualitative computer analyses of the readings detected.

Possible sources of delay here, beyond the read values from the equation of values deduced through computer output, may originate from human error, the serial interface latency, and the low-pass filter's phase shift delay[13]. For serial interference specifically, it is essential to note that overall signal latency can be affected at lower baud rates. However, the value is small enough to be minimal in terms of impact and, as such, passes the baseline value we required for the circuit to be deemed optimal.

Following Latency, the most crucial parameter discussed for this reading is the Signal-to-noise ratio. This measures the resolution of the signal detection and accounts for the circuit's efficiency in measuring and capturing a high-quality signal. Considering the applications of this device, high SNR would provide a significant benefit. However, given the experimental nature of this working paper thus far, the values can be given a certain leigh way.

---

[13] Abdullah, A. M., Ibrahim, A. R., Al-Hamadani, A. A., Al-Obaidi, M. K., & Ahmed, A. F. (2023). A review for filtering techniques of the Electrooculography (EOG) signals. *Global Journal of Engineering and Technology Advances*, *16*(3), 163–171. https://doi.org/10.30574/gjeta.2023.16.3.0177

To calculate the SNR reading for this specific circuit design, based on the parameters and output values defined through the outputted data readings, the following equation may be used:

$$\text{SNR (dB)} = 10 \times \log_{10}\left(\frac{\text{Signal Power}}{\text{Noise Power}}\right)$$

Where:

- Signal Power is the mean squared voltage: $\text{Mean}(V^2) = 5773240121$
- Noise power is the variance: $Variance(V) = 19807119$

It's essential to note the reasoning behind these two values, especially when considering the choice of statistical measurement for the power equivalent readings used in the signal-to-noise ratio equation. Signal power, the mean squared voltage, is a common interpretation of the power of a signal, and it is mainly proportional to the square of its amplitude. As such, this reading can easily define the power of the significant signal readings we must interpret and read as part of this analysis. The signal complexity makes the viable reading interpretation slightly more complex for the noise. By measuring noise power through variance, we can create a stable model that roughly predicts and estimates the differing values outputted by accurate signal shifts by the eye and actual noise that may be inherent in the circuit design. Therefore, by utilizing such an index, it's possible to capture the deviation of actual voltage readings (measured through gaze angle shift) from the mean reading value, which may be defined significantly as the values we intend to use as a baseline analysis of the readings. Variance, in simple terms, measures the degree to which the signal baseline of the mean has been "corrupted" by noise and how much deviation exists from the expected readings detected through qualitative analysis.

Applying the SNR equation to our quantified readings using Wolfram's MATHEMATICA software yields the following value:

$$\mathrm{SNR} = 10 \times \log_{10}\left(\frac{5773240121}{19807119}\right) \approx 38.65\,\mathrm{dB}$$

The signal-to-noise ratio (SNR) reading of approximately 38.65 dB demonstrates a high-fidelity signal, meaning that the valuable signal far exceeds the background noise, ensuring accurate and reliable readings. The high SNR value indicates minimal interference and distortion, suggesting that the circuit is highly efficient in transmitting clean, undistorted voltage readings. This strong separation between signal and noise ensures that the data captured is mainly free of noise-induced errors, providing confidence in the accuracy and viability of the circuit's performance. The current running theory regarding this reading is possibly mitigated interference and noise generated, and that generally, the fidelity of the TL072 chip outputted an accurate gaze angle reading. However, it is essential to note that the rate at which the signal can be calculated only relies upon the digital output this paper sought to measure. Although significant noise mitigation occurs, the amplitude of those signals reduces the ability to read accurate analog readings and, as such, limits the device to a precise, highly digital-viable reading. In contrast, the high amplitude interference and noise still cover the analog capabilities of this circuit. Although this is significant progress, it is crucial to encapsulate what areas could be improved upon considering this signal output.

## Possible significant sources of noise and interference

### Lower-than-optimal power supply reading

This issue was encountered post-testing and generally involved the less-than-optimal use of a specific voltage input for the circuit. To mitigate this next time, it is essential to use a 5.0 V signal, which is optimal for the TL072.

### Capacitive coupling and inductive interference

Electrical current from the power line generates magnetic flux that crosses the loops formed by the measurement system, inducing noise voltages at around 50 to 60 Hz.[14] The formation of inductive loops from input impedance is a highly likely contributor to the high noise in the readings. Twisting the electrode cables can quickly mitigate this.

**Baseline drift**

Movement in the eye dipole of around ±30 degrees may significantly weaken the electrode's ability to capture potential difference readings in the signal, which decreases optimal readings for the EOG circuit temporally. Errors in ocular position are the main consequence of the issue of baseline drift and contribute to an increase in noise.

**Environmental factors**

Given the high concentration of electronic devices within the room used to measure and design the circuit, the issues in reading may also have originated from interference with other device signals. This warrants the creation of a Faraday cage around the device and the shielding of the cables used with aluminum to prevent both inductive interference and device interference.

It's important to note as well, that throughout much of the experimental readings done, much uncertainty was provided as to how to properly approach the signal to noise ratio and latency detection, and the viable method encountered to do such procedure was by modeling values based on actual raw output data and utilizing that to create a model of the signal displacement from the mean values expected and the interpretation of root-mean-swuare values for voltage output, based on relevant literature on the functioning of potential difference in amplification signals and the definition of the variance of the values from the standardized mean presented.

[14] López, A., Ferrero, F., Villar, J. R., & Postolache, O. (2020). High-Performance Analog Front-End (AFE) for EOG Systems. *Electronics*, *9*(6), 970. https://doi.org/10.3390/electronics9060970

## Conclusion

This project successfully designed and developed an analog front-end for electrooculography (EOG) circuits, enabling digital communication for individuals with severe motor and communication impairments, such as those with Locked-in Syndrome (LIS) or Amyotrophic Lateral Sclerosis (ALS). Using the TL072 operational amplifier, the circuit effectively amplified weak EOG signals and interfaced with a microcontroller for real-time monitoring. Key achievements include the proper amplification and filtering of signals within the target frequency range of **0.1 Hz to 30 Hz**, allowing for accurate tracking of eye movements. The preamplifier and filtering stages reduced baseline drift and some noise, while the Arduino Uno's high sampling rate of **256 Hz** enabled reliable data collection.

The calculated Signal-to-Noise Ratio (SNR) of **38.65 dB**—well within the desired level—revealed significant efficiency in read values and the ability to measure an efficient and well-developed signal processing capability. However, it did reveal high amplitude noise interference, likely from sources such as powerline noise, capacitive coupling, and baseline drift due to electrode movement. These can be further mitigated and approached in a new way, aimed at providing a highly reliable and cost-effective method of communication utilizing the components issued in this paper. Addressing these issues will require improvements in shielding, power supply stabilization, and more effective electrode placement to minimize artifacts. Despite these obstacles, the work demonstrates the potential of low-cost analog circuits to facilitate communication for individuals with motor impairments. Future refinements could involve advanced filtering, signal optimization in interpreting the readings through a software medium, and improved electrode placement and design, further enhancing signal accuracy and usability.